\documentclass{optica-article}

\journal{opticajournal} 

\articletype{Research Article}

\usepackage{lineno}
\usepackage{siunitx}
\DeclareSIUnit\angstrom{\text {Å}}

\begin{document}

\title{Directional Dark Field for Nanoscale Full-Field Transmission X-Ray Microscopy}

\author{Sami~Wirtensohn\authormark{1,2*},  Silja~Flenner\authormark{1}, Dominik~John\authormark{1,2}, Peng~Qi\authormark{3,4}, Christian~David\authormark{3}, Manfred~May\authormark{5,6}, Patrick~Huber\authormark{5,6}, Dirk~Herzog\authormark{7,8}, Stefan~Tangl\authormark{9,10}, Carina~Kampleitner\authormark{9,10,11}, Kritika~Singh\authormark{1}, Ingomar~Kelbassa\authormark{7,8}, Katrin~Bekes\authormark{12}, Julia~Herzen\authormark{2}, Imke~Greving\authormark{1}}

\address{\authormark{1}Institute of Materials Physics, Helmholtz-Zentrum Hereon, Max-Planck-Straße 1, 21502 Geesthacht, Germany\\
\authormark{2}Research Group Biomedical Imaging Physics, Department of Physics, TUM School of Natural Sciences \& Munich Institute of Biomedical Engineering, Technical University of Munich, James-Franck-Straße 1, 85748 Garching, Germany \\
\authormark{3}Center for Photon Science, Paul Scherrer Institut, Forschungsstrasse 111, 5232 Villigen, Switzerland \\
\authormark{4}Center for Transformative Science, ShanghaiTech University, Middle Huaxia Road 393, 201210 Shanghai, China \\
\authormark{5}Institute for Materials and X-Ray Physics, Hamburg University of Technology, Denickestr. 10, 21073 Hamburg, Germany \\
\authormark{6}Center for X-Ray and Nano Science CXNS, Deutsches Elektronen-Synchrotron DESY,
Notkestr. 85, 22607 Hamburg, Germany \\
\authormark{7}Institute for Industrialization of Smart Materials, Hamburg University of Technology, Harburger Schloßstraße 28, 21079 Hamburg, Germany\\
\authormark{8}Fraunhofer IAPT, Am Schleusengraben 14, 21029 Hamburg, Germany\\
\authormark{9}Core Facility Hard Tissue and Biomaterial Research, Karl Donath Laboratory, University Clinic of Dentistry, Medical University of Vienna, Sensengasse 2a, 1090 Vienna, Austria\\
\authormark{10}Austrian Cluster for Tissue Regeneration, Donaueschingenstr. 13, 1200 Vienna, Austria\\
\authormark{11}Ludwig Boltzmann Institute for Traumatology, The Research Center in Cooperation with AUVA, Donaueschingenstr. 13, 1200 Vienna, Austria\\
\authormark{12}Department of Paediatric Dentistry, University Clinic of Dentistry, Medical University of Vienna, Sensengasse 2a, 1090 Vienna, Austria\\
}

\email{\authormark{*}sami.wirtensohn@hereon.de} 

\begin{abstract*}
Dark-field X-ray imaging visualizes structural inhomogeneities through small-angle scattering, but existing directional methods are confined to the micrometer scale. While recent advances have extended dark-field capabilities to nanoscale transmission X-ray microscopy (TXM), directional scattering retrieval -- critical for characterizing anisotropic nanostructures -- has remained inaccessible for imaging resolutions in the sub-micrometer scale.
Here, we demonstrate the first directional dark-field setup for nanoimaging, achieving orientation mapping of scattering features below the spatial resolution limit. Our method is experimentally simple to implement with existing TXM setups. We validate its performance by successfully resolving sub-resolution test structure orientations, cross-correlating orientational changes within hierarchical nanoporous materials, and mapping the directional arrangement of hydroxyapatite nanocrystals (30-\SI{70}{\nano\meter}) within human tooth enamel. By utilizing shadow regions in the optical configuration, we further extend the detectable scattering vector range, demonstrating a pathway toward size-selective dark-field imaging. This advancement enables the quantitative structural characterization of anisotropic nanomaterials, which are critical to biomineralization, advanced materials, and nanotechnology applications.
\end{abstract*}

\section{Introduction}
Directional dark-field imaging is an emerging technique that enables access to the preferential scattering direction of materials \cite{jensen2010directional, smith2022x}. However, until now, the implementation of directional dark-field imaging has been limited to imaging in the micrometer regime \cite{croughan2023directional, Lautizi2025}. Here, we extend directional dark-field imaging to the nanoscale for the first time, enabling the investigation of oriented nanostructures in hierarchical materials with sub-micrometer resolution.

The ability to visualize and characterize the hierarchical 3D structure and inhomogeneities of materials is crucial across a range of scientific fields like medicine~\cite{walsh2021imaging,besnard2023synchrotron}, biology \cite{albers2024synchrotron}, and materials science~\cite{lavery2014x, Gries2023}. A popular method for investigating structural information in real space is transmission X-ray microscopy (TXM) \cite{andrews2011transmission, spence2021transmission}, which utilizes the sample's attenuation as a contrast mechanism. However, the limitations of attenuation-based X-ray imaging in detecting subtle sample features have driven the development of complementary imaging modalities such as phase-contrast imaging, which exploits the refractive properties of a sample to enhance image contrast~\cite{Zernike1942, Schmahl1994, Neuhausler2003, Stampanoni2010, Flenner2023}. 

Full-field dark-field imaging is another approach, which is based on the small-angle X-ray scattered (SAXS) intensity rather than attenuation or phase, and therefore provides a unique contrast mechanism that highlights structural inhomogeneities invisible to conventional techniques \cite{Rigon2007, Pfeiffer2008, Wirtensohn2024}. 
Dark-field imaging has proven to be a valuable tool to gain deeper insights into the sample structure \cite{Jud2021}. In medical research, it aids in detecting and quantifying lung diseases like emphysema \cite{hellbach2015vivo, urban2022qualitative} and improving breast cancer diagnostics by revealing the micromorphology of breast calcifications \cite{grandl2015improved, arboleda2020towards}. In manufacturing, it identifies defects in glass fiber-reinforced thermoplastics \cite{Ozturk2023}, visualizes flaws in composite structures \cite{Endrizzi2015}, and detects porosity below pixel resolution \cite{Revol2011}. 

Over the past years, multiple approaches have been developed to recover the dark-field signal. In analyzer-based methods, a crystal is used to reflect only X-rays fulfilling the Bragg condition. By rocking the crystal over a range of angles, the broadening of the X-ray beam due to scattering within the sample can be measured and used to retrieve the full-field dark-field image \cite{Rigon2007}. In reference pattern-based methods, a wavefront marker is introduced into the beam path; the dark-field signal is then extracted by tracking the sample-induced changes to the pattern either locally \cite{Morgan2011, Berujon2012, zdora2017x, zdora2018state} or globally \cite{Paganin2018, Pavlov2020, alloo2022dark, Beltran2023}. Examples of direct imaging methods used for the dark field are single-grid \cite{How2022, How2023} and speckle-based imaging \cite{zdora2018state, Savatovic2024}, as well as beam-tracking edge illumination \cite{Dreier2020}. If the reference pattern is not directly resolvable, the dark-field signal can also be extracted indirectly by employing multiple scanning steps, as utilized in Talbot-Lau interferometry \cite{David2002, Pfeiffer2008, pfeiffer2009x} and edge illumination \cite{Endrizzi2014, Ozturk2023}. Recent developments also allow the optics-free extraction of the dark-field signal based on the Fokker-Planck equation by utilizing multiple distances \cite{leatham2024x} or multiple energies \cite{ahlers2024x}. Most dark-field methods are currently limited to micrometer resolution. Only recently, it has been shown for the first time that dark-field imaging can be extended to the nanoscale based on a full-field TXM approach \cite{Wirtensohn2024}.

While dark-field imaging detects the overall scattered intensity per pixel, additional information can be uncovered in the case of samples with structured features, which create anisotropic scattering. Aligned sample features with a high orientation dependence scatter strongly perpendicular to that orientation and weakly in parallel directions. This asymmetric scattering signal can be exploited to visualize the orientation of the scattering features within the sample, as recently demonstrated in visible light microscopy by an aperture scanning approach~\cite{zheng2025optical}. In X-ray full-field imaging, the visualization of the orientations is called directional dark-field imaging \cite{jensen2010directional, croughan2023directional} and has recently been used to investigate the microstructure in archaeological skeletal remains \cite{Lautizi2025} and the orientation within carbon fiber reinforced polymers \cite{smith2022x}. The directional characteristic, combined with the high sensitivity of the dark field to small internal structures, such as cracks, bubbles, and material boundaries, makes it indispensable for understanding material properties in materials science. However, until now, the implementation of directional dark-field imaging is limited to the micrometre regime.

This study introduces the first successful demonstration of directional dark-field imaging at the nanoscale in a full-field TXM, enabling the orientation mapping of sub-resolution features by adding apertures in front of the condenser.
We show the abilities of the directional dark-field TXM on a Siemens star test object, a hierarchical nanoporous silicon pillar, and the enamel of a human permanent tooth. Additionally, we extend the maximum magnitude of the scattering vector by manipulating the illumination function to enhance the detectability of smaller feature sizes. These advancements in nanoscale dark-field imaging open up new possibilities for gaining additional insights into material properties across multiple scientific disciplines and allow the extraction of structural information even in projection space, ideal for high-resolution, multimodal imaging applications.

\section{Method}
The experiments were conducted at the PETRA~III P05 nanotomography endstation at DESY in Hamburg, Germany. The P05 beamline is operated by the Helmholtz-Zentrum Hereon. A 2-meter-long U29 undulator generates the X-ray beam and produces a source size of \SI{36.0}{\micro\meter}~×~\SI{6.1}{\micro\meter} with a divergence of \SI{28.0}{\micro\radian}~×~\SI{4.0}{\micro\radian}. A Si-111 double-crystal monochromator tunes the beam to an energy of \SI{11}{\kilo\electronvolt}. The imaging is performed using a Hamamatsu C12849-101U detector, featuring a pixel size of \SI{6.5}{\micro\meter}, a 2048 × 2048 pixel array with a 1:1 fiber coupled \SI{10}{\micro\meter} thick Gadox scintillator \cite{Flenner2020, Flenner2022p}. The detector captures 16-bit images and is located in the adjacent experimental hutch, roughly \SI{19}{\meter} downstream from the sample. During the experiments, PETRA~III was operated in multi-bunch mode with a beam current of \SI{100}{\milli\ampere}, except for the hierarchical nanoporous silicon, which was conducted with a beam current of \SI{120}{\milli\ampere}.

\subsection{Dark-field transmission X-ray microscopy}
\begin{figure}[htbp]
    \centering\includegraphics[width = \textwidth]{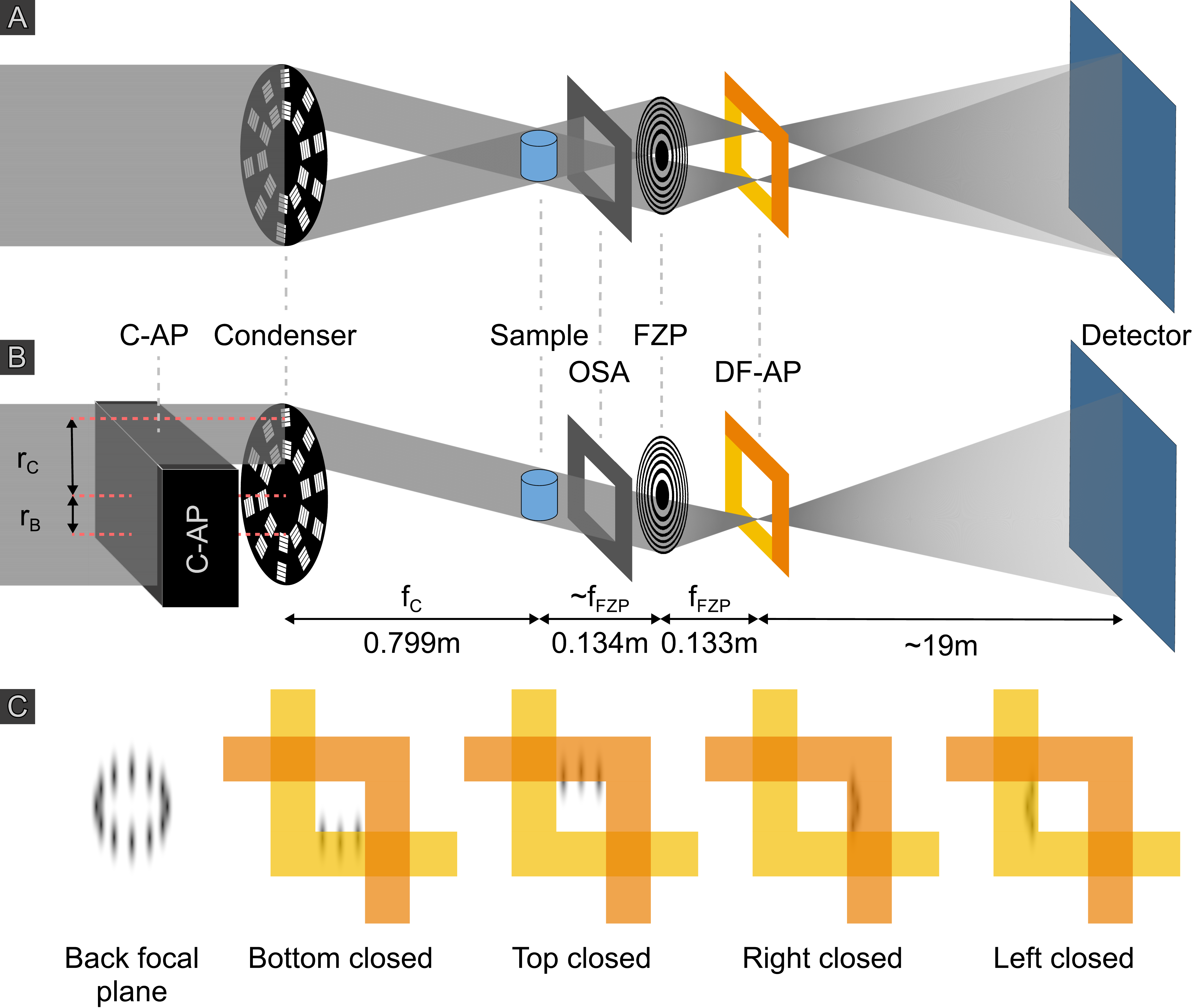}
     \caption{Schematic of the normal dark-field TXM setup (A). The beam shaping condenser splits the beam into multiple deflected parallel beams, creating a ring of focused points in the back focal plane of the Fresnel zone plate (FZP), which can be blocked by dark-field apertures (DF-AP) to only let the scattered light pass to the detector. With an additional condenser aperture (C-AP), the setup can be extended to directional dark-field imaging (B). The C-AP blocks two-thirds of the condenser fields and only allows light from one direction to illuminate the sample. For non-uniform scattering structures, the focal spots are elongated (C). By successively closing the C-AP in the four directions (bottom, top, right, left), the orientation of the scattering can be retrieved. Here, visualized for scattering in the vertical direction. (A) is adapted with permission from \cite{Wirtensohn2024} \textcopyright Optica Publishing Group.}
    \label{fig:setup}
\end{figure}

The dark-field transmission X-ray microscope used for the experiments is based on the design of Wirtensohn et al. and consists of a beam shaping condenser, a Fresnel zone plate (FZP) and a dark-field aperture (DF-AP) \cite{Wirtensohn2024}. As illustrated in Figure \ref{fig:setup} A, the condenser splits the incoming parallel beam into multiple square-shaped beamlets and redirects them by diffraction onto the sample, creating a flat-top illumination with a size of \SI{100}{\micro\meter} x \SI{100}{\micro\meter} \cite{Jefimovs2008, Vartiainen2014}. Each beamlet in itself still has the properties of a parallel beam. After the interaction with the sample, the beamlets are focused by the FZP, serving as an objective lens, and a ring of focal spots is created in the back focal plane of the FZP. This ring of focal spots encloses the shadow area originating from the beam stop. As the focal spots propagate, they expand in size and finally overlap to create a single image of the sample in the detector plane.

If the sample contains scattering features, the light from the focal spots is broadened into the shadow area within the ring of focal spots. By using two L-shaped apertures in the back focal plane of the FZP, the focal spots can be blocked and only the light cast into the shadow area can reach the detector. This produces a dark-field image that contains only information about the scattering characteristics of a sample \cite{Wirtensohn2024}.

\subsection{Directional dark field}
The shape of a single focal spot in the back focal plane of the FZP can be approximated as Gaussian-like. This Gaussian shape has a specific width and is entirely obstructed by the DF-AP in the absence of scattering. Any scattering of a sample broadens the peak. This peak broadening can be uniform or have a specific orientation, depending on the structural features of the sample. In all cases, the contribution of each focal spot to the dark-field image is solely the portion that is diffracted into the shadow region.

The contributing part of a focal spot to the dark-field signal depends on the direction from which the beam comes and from which direction the focal spot is blocked: If the beam comes from the top of the condenser, the bottom DF-AP blocks the signal, and only the upper part of the scattered signal can pass. Conversely, if the beam comes from the right part of the condenser, the left DF-AP blocks the signal, and only the right part of the scattered signal can pass. Since the condenser creates beams from various directions, the signal is equally sensitive to all directions. However, we can utilize this to influence the directions contributing to the dark-field signal by partially covering the condenser and allowing only specific beams with a given direction to pass. To achieve this, we introduce two further L-shaped apertures upstream of the condenser. This condenser aperture (C-AP) is placed on piezoelectric actuators and allows fine movements to block the square-shaped gratings of the condenser, which create the individual focal spots in the back focal plane of the FZP. This way, we can select the direction we want to be sensitive to, as shown in Figure \ref{fig:setup} B, where the C-AP covers the bottom of the condenser, allowing only beams from the top of the condenser to illuminate the sample. In this example, the corresponding focal spots are located in the bottom part of the back focal plane, and if blocked by the DF-AP, only the upper part of the focal spots will contribute to the dark-field signal.

If a perfectly uniformly scattering sample is considered, the Gaussian shape of a focal spot is broadened in all directions equally. This means the same signal will be detected when covering the bottom, top, right, or left side of the condenser with the C-AP, since the focal spots contribute in the same way to the dark-field signal.

If, on the other hand, the sample scatters more in one specific direction, the shape of the Gaussian peak becomes elongated in that direction. Consequently, the focal spots do not contribute in the same way anymore if the C-AP covers different parts of the condenser, as shown in Figure \ref{fig:setup} C. Assume a sample that only scatters in the vertical direction. By utilizing only the light from the top of the condenser, some of the scattered signal can pass the DF-AP (Figure \ref{fig:setup} C „Bottom closed“) and therefore the scattering structure will be visible on the detector. However, if the light from the left of the condenser is used, none of the light is scattered into the shadow area, and the structure is therefore not visible on the detector (Figure \ref{fig:setup} C „Right closed“). The C-AP can hence be used to create a direction-selective dark-field image, which only contains scattering information in one direction.

To retrieve the directionality of the dark field, four projections are taken. For each projection, the C-AP covers two-thirds of the condenser from a different direction (bottom, top, left, right) as shown in Figure \ref{fig:setup} C. Blocking two-thirds optimizes the direction-selectivity and the flux. Each projection $I$ is dark current, bad pixel, and beam current corrected. Subsequently, the sample projection $I_{\text{s}}$ is subtracted by the corresponding background scattering intensity $I_{\text{b}}$. The background scattering is retrieved for each C-AP position by taking a projection with closed DF-AP and without the sample in the beam. Considering the dark-field image with the right side of the condenser covered by the C-AP, $D_{\text{r}}$, is given by
\begin{align}
    D_{\text{r}} = I_{\text{r, s}} - I_{\text{r, b}},
    \label{eq:Dr}
\end{align}
with $I_{\text{r, s}}$ and $I_{\text{r, b}}$ being the projections of the sample and the background with the right side of the condenser covered. A pair of projections is combined to create the dark-field image of one direction
\begin{align}
\label{eq:Dx}
    D_{\text{x}} = \frac{D_{\text{r}} + D_{\text{l}}}{2}, \\
\label{eq:Dy}
    D_{\text{y}} = \frac{D_{\text{t}} + D_{\text{b}}}{2},
\end{align}
with $D_{\text{x}}$ being the dark-field image in x-direction and $D_{\text{y}}$ in y-direction. $D_{\text{l}}$, $D_{\text{t}}$, and $D_{\text{b}}$ are the dark-field signals where the C-AP covers correspondingly the left, top, and bottom part of the condenser. Based on $D_{\text{x}}$ and $D_{\text{y}}$ an angular map with a \SI{90}{\degree} range is created
\begin{align}
    \Phi = \text{atan2}(D_{\text{x}}, D_{\text{y}}),
    \label{eq:angle}
\end{align}
as well as a magnitude map
\begin{align}
    M = \sqrt{D_{\text{x}}^2 + D_{\text{y}}^2}.
    \label{eq:magnitude}
\end{align}
The two maps are combined to create a composite image in which the angular map $\Phi$ determines the color and the magnitude map $M$ its luminance. The resulting image is further referred to as the directional dark field. The magnitude map can therefore also be understood as a confidence map for the retrieved angles.

\subsection{Extension of the scattering vector}
The smallest size of the scattering structures detectable by this dark-field setup is related to the maximum magnitude of the scattering vector $|\Vec{Q}_{\text{max}}|$ which is given by
\begin{align}
    |\Vec{Q}_{\text{max}}| = \frac{\pi}{\Delta r_{\text{C}}} \left(\frac{r_{\text{B}}}{r_{\text{C}}} + 1 \right),
\label{eq:Q_max_f}
\end{align}
with $r_{\text{C}}$ being the radius of the condenser, $r_{\text{B}}$ the radius of the beamstop, and $\Delta r_{\text{C}}$ the outermost zone width of the condenser \cite{Wirtensohn2024}. Given an identical optics arrangement, only $r_{\text{B}}$ can be increased to push towards smaller structure sizes. However, since more of the condenser area is covered by the beamstop, the flux will decrease by a factor based on the covered area.

\begin{figure}[htbp]
    \centering\includegraphics[width = \textwidth]{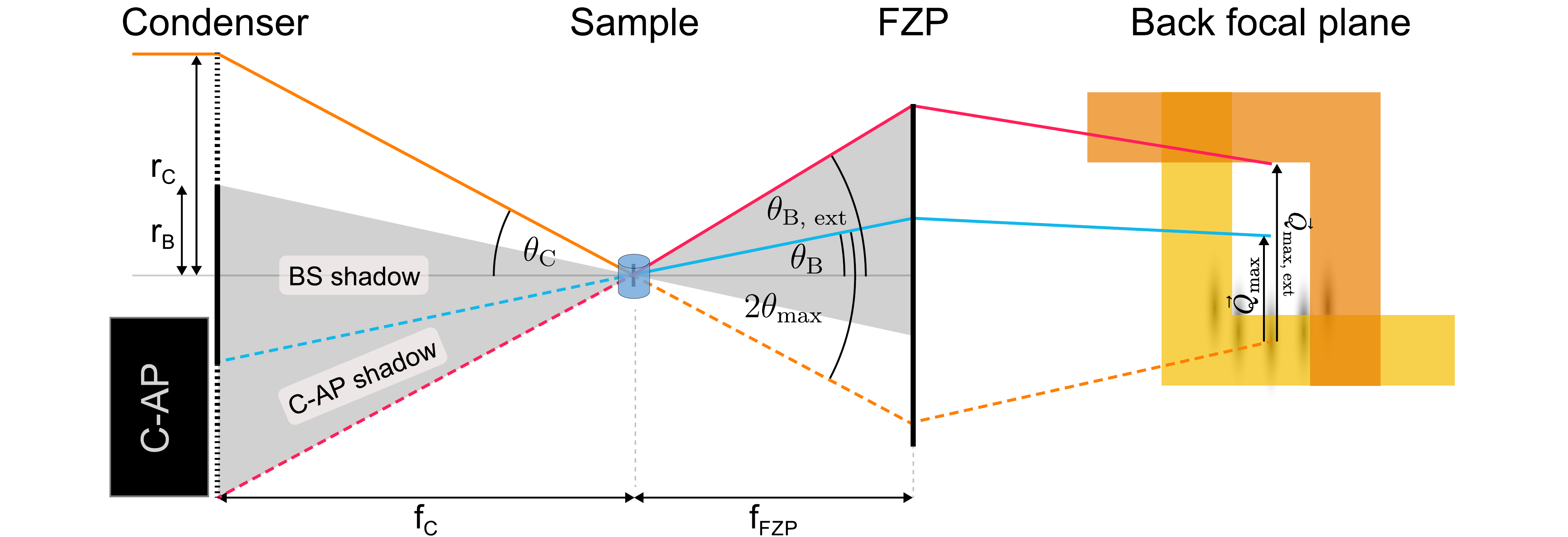}
    \caption{Extension of the maximal magnitude of the scattering vector $|\Vec{Q}_{\text{max, ext}}|$. By covering two-thirds of the condenser, the shadow area (gray) is elongated towards the opposite side. This allows further opening of the dark-field apertures in the back focal plane, extending $|\Vec{Q}_{\text{max, ext}}|$. The maximal scattering angle is achieved when the outermost beamlet from the condenser (orange solid line) gets scattered towards the edge of the shadow area (pink solid line). By using the additional shadow area, the scattering angle increases from the blue solid line to the pink solid line.}
    \label{fig:df_extension}
\end{figure}

For the measurements of the directional dark-field image, two-thirds of the condenser is always covered by the C-AP. This covered area creates an enlarged shadow in the back focal plane of the FZP as shown in Figure \ref{fig:df_extension}. The elongated shadow towards the opposite side of the C-AP can be used to extend the maximum scattering vector by opening the DF-AP in the corresponding direction. This enables the detection of smaller scattering structures. In doing so, the magnitude of the maximum scattering vector is limited by the radius of the FZP. This is because the further scattered intensity will miss the FZP and therefore not contribute to the dark-field image formation.

To calculate the extended maximal magnitude of the scattering vector $|\Vec{Q}_{\text{max, ext}}|$ for the directional dark-field setup, the magnitude of the scattering vector $|\Vec{Q}|$ with the scattering angle $\theta$ is considered
\begin{align}
    \label{eq:Q}
    |\Vec{Q}| = \frac{4\pi \sin{\theta}}{\lambda},
\end{align}
with $\lambda$ being the energy dependent wavelength.
As shown in Figure \ref{fig:df_extension}, the scattering vector becomes maximal for the directional dark-field setup if the outermost beamlet coming from the condenser (orange solid line) with an angle $\theta_{\text{C}}$ is scattered towards the end of the shadow area at the edge of the FZP (pink solid line) with the angle $\theta_{\text{B, ext}}$. Therefore the maximum of the extended scattering angle $\theta_{\text{max, ext}}$ is given by 
\begin{align}
    \label{eq:theta_max_ext}
    2\theta_{\text{max, ext}} = \theta_{\text{C}} + \theta_{\text{B, ext}}.
\end{align}
Based on geometrical considerations, the opening angles are
\begin{align}
    \label{eq:thetaCthetaB}
    \theta_{\text{C}} \approx \tan{\theta_{\text{C}}} = \frac{r_{\text{C}}}{f_{\text{C}}}, && \theta_{\text{B, ext}} \approx \tan{\theta_{\text{B, ext}}} = \frac{r_{\text{FZP}}}{f_{\text{FZP}}},
\end{align}
with $r_{\text{FZP}}$ the radius and $f_{\text{FZP}}$ the focal distance of the FZP and $f_{\text{C}}$ the focal distance of the condenser. By combining equation~\eqref{eq:theta_max_ext} and equation~\eqref{eq:thetaCthetaB} and inserting them into equation \eqref{eq:Q} the extended maximal magnitude of the scattering vector becomes
\begin{align}
    |\Vec{Q}_{\text{max, ext}}| = \frac{2 \pi}{\lambda} \left(\frac{r_{\text{C}}}{f_{\text{C}}} + \frac{r_{\text{FZP}}}{f_{\text{FZP}}}\right).
\end{align}
If the numerical aperture of the BS and the FZP is the same, $\frac{r_{\text{FZP}}}{f_{\text{FZP}}}$ can be substituted by $\frac{r_{\text{C}}}{f_{\text{C}}}$ resulting in
\begin{align}
    \label{eq:Q_rc}
    |\Vec{Q}_{\text{max, ext}}| = \frac{4 \pi}{\lambda} \frac{r_{\text{C}}}{f_{\text{C}}}.
\end{align}
The focal distance of the condenser $f_{\text{C}}$ is defined as
\begin{align}
    f_{\text{C}} = \frac{2 r_{\text{C}} \Delta r_{\text{C}}}{\lambda},
\end{align}
and put into equation~\eqref{eq:Q_rc} yields
\begin{align}
    |\Vec{Q}_{\text{max, ext}}| = \frac{4 \pi}{\lambda} \frac{r_{\text{C}} \lambda}{2r_{\text{C}} \Delta r_{\text{C}}} = \frac{2 \pi}{\Delta r_{\text{C}}}.
\end{align}
This shows that the maximal magnitude of the extended scattering vector only depends on the outermost zone width of the condenser if both optics have the same numerical aperture. The smallest scattering features $d_{\text{ext}}$ of the extended directional dark-field setup can then be estimated by 
\begin{align}
    d_{\text{ext}} = \frac{2 \pi}{|\Vec{Q}_{\text{max, ext}}|} = \Delta r_{\text{C}}.
\end{align}
It should be noted that this extension of the scattering vector is only towards one direction, the direction covered by the C-AP.

\section{Experimental results}

\subsection{Directional dark field}
\begin{figure}[htbp]
    \centering\includegraphics[width = \textwidth]{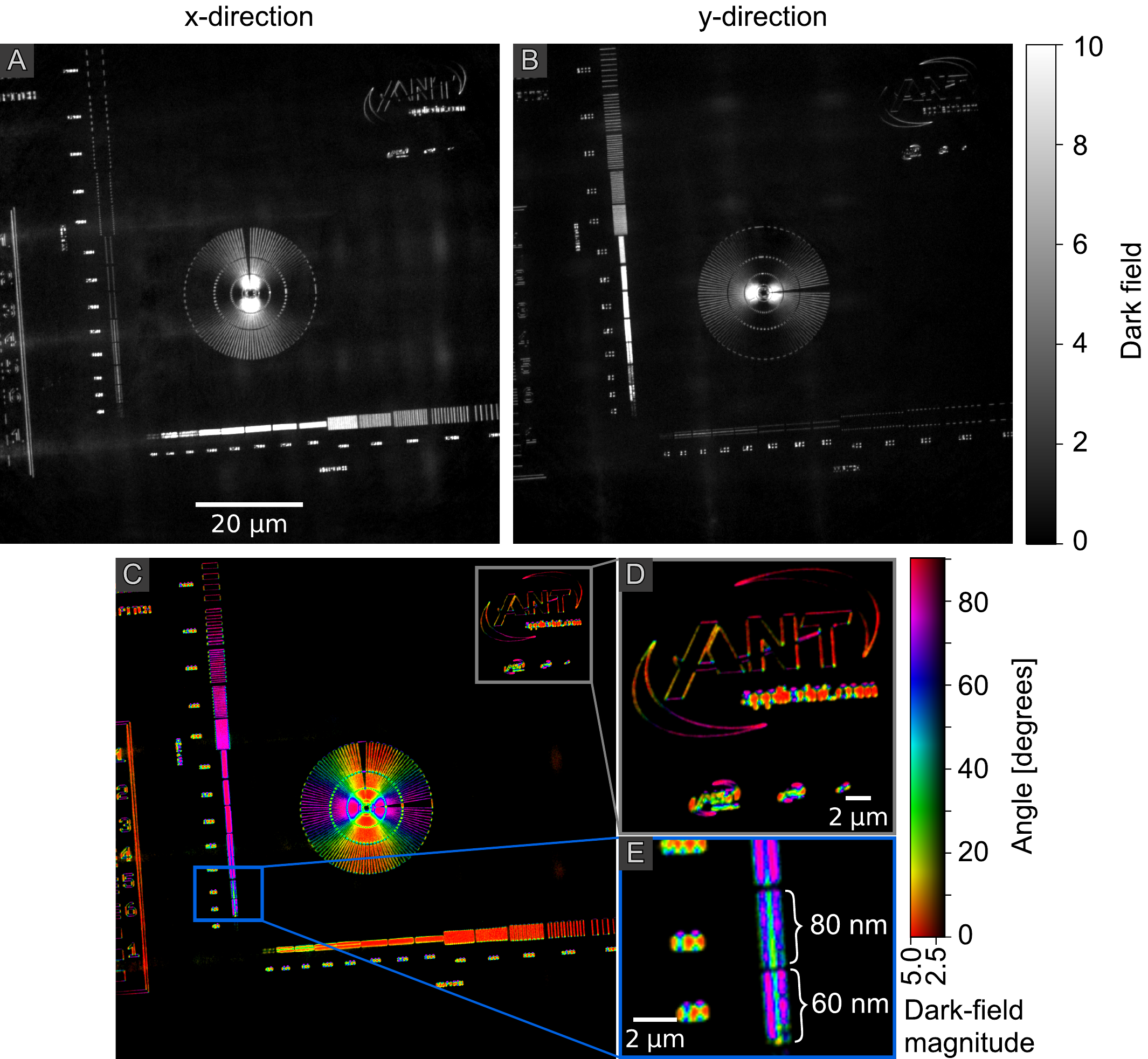}
    \caption{Directional dark field of a Siemens star. The directional dark-field components in the x- and y- directions~(A and B), originating from the C-AP being closed in left and right~(A) and top and bottom~(B), are used to calculate a scattering vector. The angle and magnitude of the resulting vector are used to create a color plot~(C). The color corresponds to the scattering angle with respect to the vertical axis, and the luminance to the magnitude of the scattering vector. The different directions can be well separated~(D), and works also for structure sizes below the spatial resolution of the setup, as visible for the line pairs with a pitch of \SI{80}{\nano\meter} (\SI{40}{\nano\meter} feature size) and \SI{60}{\nano\meter} (\SI{30}{\nano\meter} feature size)~(E). The total exposure time of the directional dark-field image is $4 \times \SI{300}{\second} = \SI{20}{\minute}$.}
    \label{fig:direc_df}
\end{figure}

As a test pattern, a golden Siemens star with structure heights of \SI{600}{\nano\meter} is used. On the same membrane, nearly horizontally and vertically oriented line patterns are located with decreasing structural sizes towards the bottom left. By closing the DF-AP the TXM is switched to the normal dark-field modality. By covering the condenser with the C-AP the setup becomes sensitive to the orientation of the dark-field signal. By combining the projections taken with the C-AP closed once on the left side and once on the right side, the signal in the x-direction can be gained based on equation~\eqref{eq:Dx} as shown in Figure \eqref{fig:direc_df} A. Here, only structures running in the vertical direction are visible, while the structures running in the horizontal direction do not contribute to the image. This is the case since edges create a SAXS signal mainly perpendicular to their longitudinal direction. In Figure \ref{fig:direc_df} A, this is evident when looking at the line structures at the bottom with vertical orientation and the left side with horizontal orientation. While the structures at the bottom light up, the structures on the left are not visible. The only part apparent of the left line pairs are the side edges of the lines, which again create a signal in the x-direction.
The Siemens star in Figure \ref{fig:direc_df} A also shows a strong angular dependency of the dark-field signal. The intensity gradually decreases from vertical to horizontal structures, illustrating the directional dependency of the signal. Figure \ref{fig:direc_df} B shows the same test pattern. However, it was created by combining the projections with the C-AP closed, once on the top side and once on the bottom side, and therefore showing the dark-field signal in y-direction.
The direction-selective dark-field projections in x- and y-direction are used to calculate, based on equation~\eqref{eq:angle} and \ref{eq:magnitude}, the directional dark-field image shown in Figure \ref{fig:direc_df} C. Here, the color map shows the angular orientation of the structures with respect to the vertical axis, while the brightness visualizes the scattering magnitude of the structures. The directional dark-field image immediately reveals the orientation of the scattering structures as visible at the Siemens star. The orientation of smaller structures can also be well separated as shown in Figure \ref{fig:direc_df} D, where the outer circular parts of the logo show a continuous change in color and, therefore, of the scattering orientation. This is also true for feature sizes below the spatial resolution of the setup. In such a case, the directional dark field does not directly resolve these smaller features, but since they are still scattering, they can be visualized, and their average orientation within one pixel can be retrieved. The spatial location remains limited to the spatial resolution of the setup. Figure \ref{fig:direc_df}~E shows line pairs with a pitch of \SI{60}{\nano\meter} (\SI{30}{\nano\meter} feature size). Based on the directional dark-field image, the sub-resolution features can not only be visualized, but also information about their orientation can be gained. Some parts of the line pairs show inconsistencies within their orientation, leading to the conclusion that these lines are collapsed. Also, note that the text to the left of the line pairs can be separated into different directions, even though it is not fully resolved and readable.

\begin{figure}[htbp]
    \centering\includegraphics[width = \textwidth]{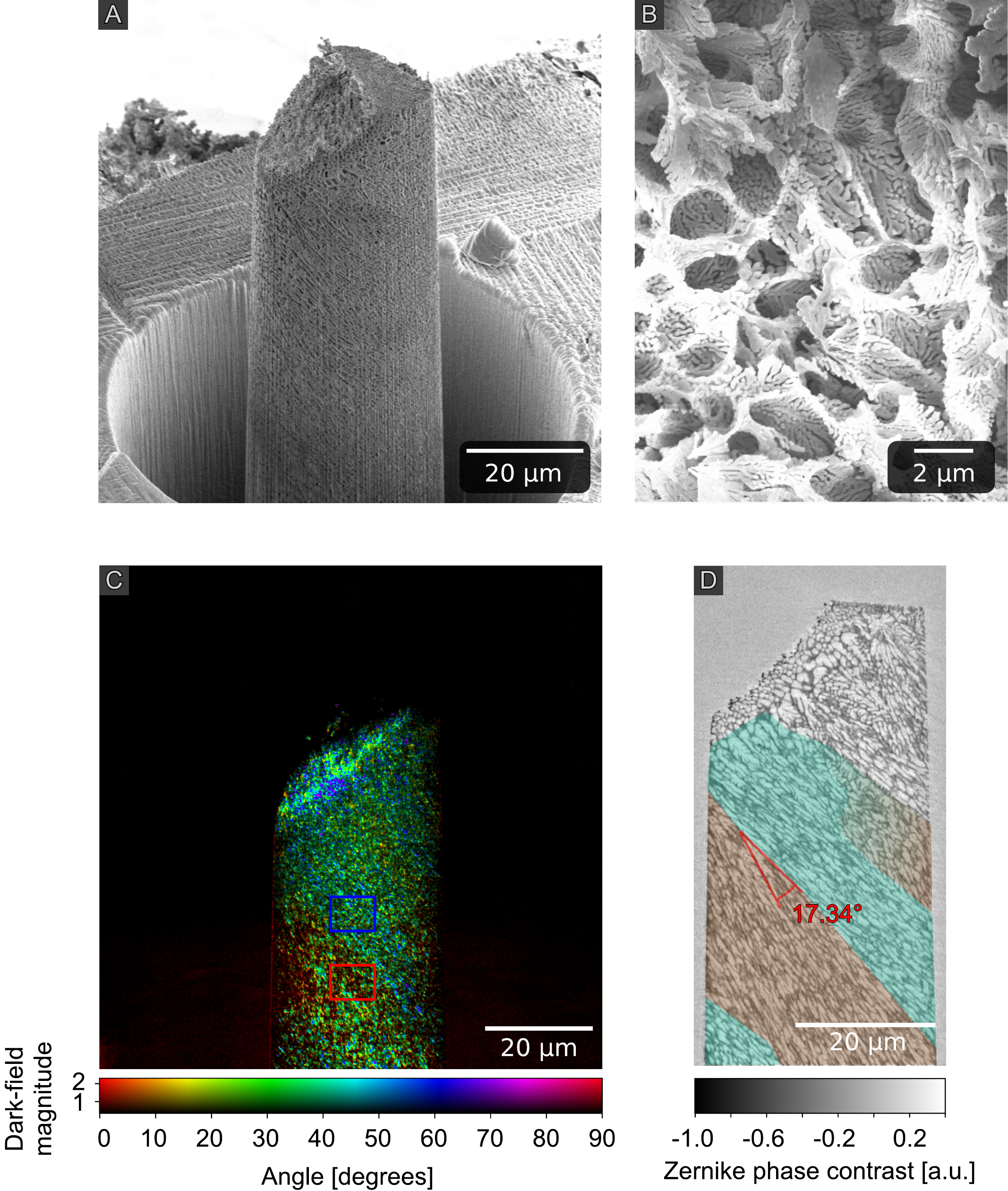}
    \caption{SEM image of a highly porous hierarchical silicon pillar~(A), which consists of \SI{1}{\micro\meter} to \SI{8}{\micro\meter} large, elongated pores. These pores consist of individual ligaments ranging from \SI{50}{\nano\meter} to \SI{200}{\nano\meter} in diameter. Their alignment along the major axis of the pores creates an anisotropic dark-field signal~(B). The directional dark-field projection of the sample reveals directional changes in the internal structure of the pillar. The difference of the magnitude-weighted mean directional dark-field angle in the blue and red region of interest is $\SI{18.72}{\degree} \pm \SI{0.28}{\degree}$~(C). A vertical slice through a Zernike phase contrast measurement validates an orientational change in the sample's pores of approximately \SI{17.34}{\degree}, visualized by the colored areas~(D). The total exposure time of the directional dark-field image is $4 \times \SI{250}{\second} = \SI{16}{\minute} \ \SI{40}{\second}.$}
    \label{fig:si_direc_df}
\end{figure}

As a second sample, a hierarchical nanoporous silicon pillar is measured. The material is created through additive manufacturing via powder bed fusion by laser beam (PBF-LB/M) of an aluminium-silicon alloy and a subsequent dealloying step~\cite{PhilippTimm2023}. Due to the rapid solidification process during PBF-LB/M and depending on the heat flow direction, an anisotropic material emerges. In the as-built state, silicon forms a thin interconnected network within the aluminium matrix. After dealloying, only the silicon remains, creating a highly porous material as shown in Figure~\ref{fig:si_direc_df}~A. The sample used here features a porosity of around 89~\%. Its structure consists of large, elongated pores with sizes in the range of \SI{1}{\micro\meter} to \SI{8}{\micro\meter}. As shown in Figure~\ref{fig:si_direc_df}~B, these pores are composed of individual ligaments ranging from \SI{50}{\nano\meter} to \SI{200}{\nano\meter} in diameter. The orientational preference of the pores and nanostructures creates a structured dark-field signal.
The directional dark-field projection in Figure~\ref{fig:si_direc_df}~C reveals orientational changes within the sample. In the central region of the pillar, two differentiable areas are discernible, which run diagonally from the top left to the bottom right. The upper region displays more green and blue colors, compared to the lower region, which features more red and yellow colors. This indicates an internal directional change of the structures. Since the magnitude can be understood as a confidence map for the directional dark-field angles, it enables a quantitative comparison of the magnitude-weighted mean angle between two regions of interest. Looking at the magnitude-weighted mean in the mentioned regions, marked by the blue and red rectangles in Figure~\ref{fig:si_direc_df}~C, the measured angular difference is $\SI{18.72}{\degree} \pm \SI{0.28}{\degree}$.
Figure~\ref{fig:si_direc_df}~D shows the central vertical slice through the same pillar measured via Zernike phase contrast (ZPC) TXM. In the ZPC slice, the pores as well as their elongated shapes are clearly visible. Examining the orientation of the pores reveals two main regions marked by light green and red color. These regions are in good agreement with the orientations of the angular map from the directional dark-field projection. Measuring the angle between the main orientation of these regions in the ZPC slice yields a difference of \SI{17.34}{\degree}, which matches the measurement of the directional dark-field signal.
We note here that the dark field is a projection, which provides the average angular orientation throughout the entire sample thickness at a given pixel position, whereas the ZPC slice only shows the local change of the sample at the specified depth. Hence, a slight difference naturally occurs between the two measurements.

\begin{figure}[htbp]
    \centering\includegraphics[width = \textwidth]{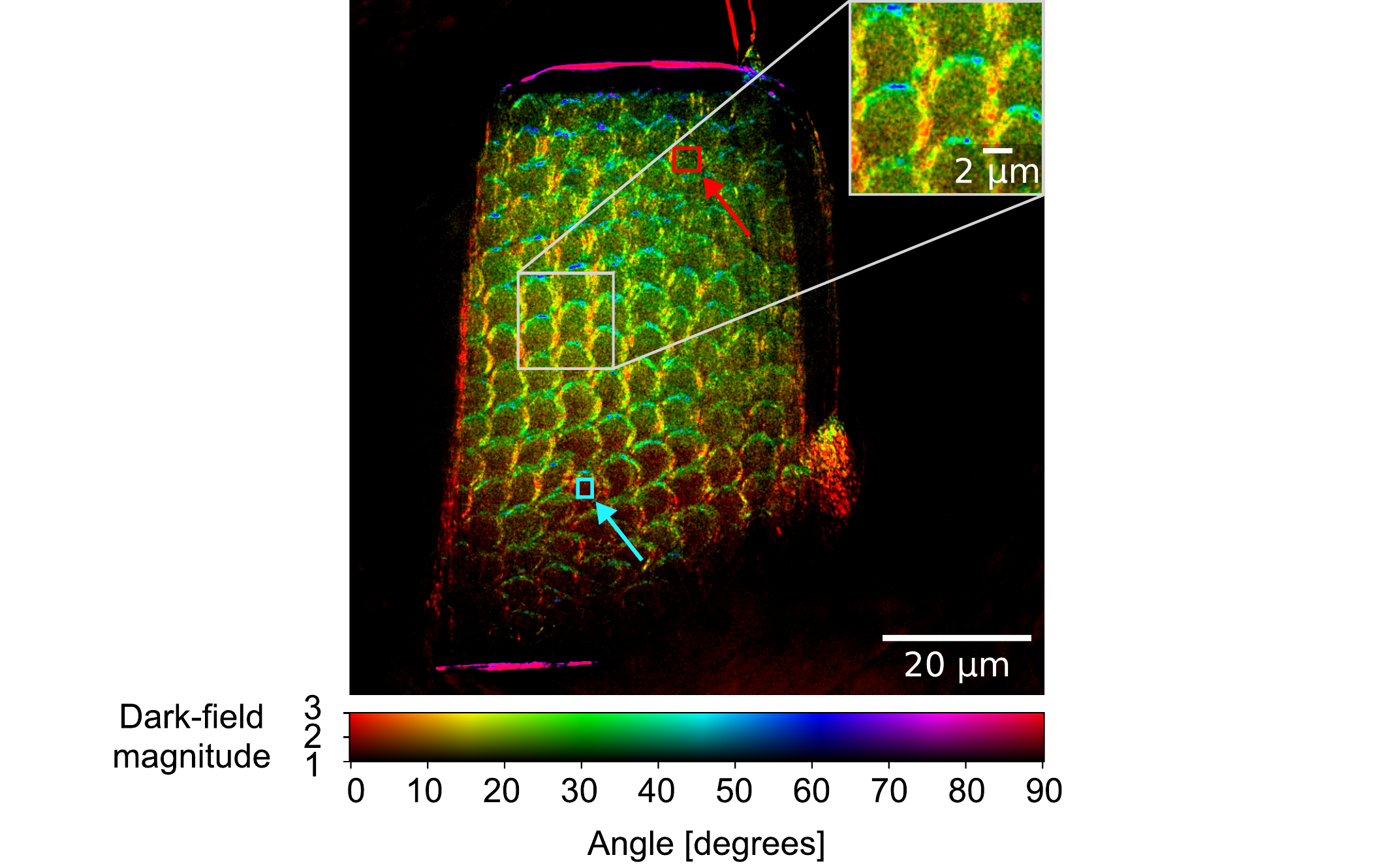}
    \caption{Directional dark-field projection of the enamel of a human permanent MIH-tooth. The enamel consists of hydroxyapatite crystals (\SI{30}{nm} - \SI{70}{nm}), which are bundled into rods (around \SI{6}{\micro\meter}). These are visible as fish-scale-like structures. Due to the crystalline structure, the rods have a strong directional dependency, which can be visualized by the directional dark field. The difference in the magnitude-weighted angle of the directional dark-field image between the top right (red) and bottom left (turquoise) region of interest is $\SI{22.23}{\degree} \pm \SI{0.28}{\degree}$. The total exposure time of the directional dark-field image is $4 \times \SI{300}{\second} = \SI{20}{\minute}$.}
    \label{fig:tooth_direc_df}
\end{figure}

As a third sample, a pillar from the enamel of a human permanent tooth with molar incisor hypomineralization (MIH) is imaged. The condition is defined as hypomineralization of the enamel of at least one first permanent molar with or without the involvement of incisors. It is a prevalent condition affecting children worldwide~\cite{Schwendicke2018}. The enamel of a human tooth consists mainly of hydroxyapatite crystals \cite{Robinson2004}. They have a width of \SI{30}{\nano\meter} to \SI{70}{\nano\meter}, and a length of \SI{100}{\nano\meter} to \SI{1000}{\nano\meter}~\cite{Wang2006}. The crystals bundle into rods, which are about \SI{6}{\micro\meter} in diameter and are also referred to as prisms~\cite{Fosse1968}. The crystals and the edges of the prisms create a structured dark-field signal. Figure~\ref{fig:tooth_direc_df} shows the directional dark-field projection of the tooth sample. The outer edges of the prisms dominate the dark-field image. Their orientation can be well separated, as seen in the zoom-in region. Compared to healthy enamel, a more marked inter-prismatic space can be found in the examined sample. This is a typical pattern that can be seen in MIH-affected enamel~\cite{Fagrell2010}. Besides the edges, there is also a dark-field signal coming from within the prisms. This signal originates from the crystalline structure of the hydroxyapatite. Note how the orientation angle, and therefore the color, changes from the bottom left to the top right of the sample. The change in color could be related to the change in the orientation of the crystals within a prism. Figure \ref{fig:tooth_direc_df} shows the prisms as circular structures with an opening on one side, known as a keyhole structure. In the lower-left region of the tooth, the openings are oriented approximately \SI{45}{\degree} downwards-left. Toward the top-right region of the tooth, these openings gradually rotate and become increasingly oriented downwards. This shift in the orientation of the keyhole structure also affects the orientation of the crystals within. When comparing a small region within a keyhole at the top right side (red region of interest) and a region within a keyhole at the bottom left side (turquoise region of interest), an orientational change of $\SI{22.23}{\degree} \pm \SI{0.28}{\degree}$ can be measured based on the difference in the magnitude-weighted mean angle.
This is a strong indication that the directional dark-field signal coming from within the keyhole structures is sensitive to the change of orientation of the hydroxyapatite crystals. Furthermore, note that the solid support structures on the top and right side are completely dark and do, therefore, not contain any scattering structures.

\begin{figure}[htbp]
    \centering\includegraphics[width = \textwidth]{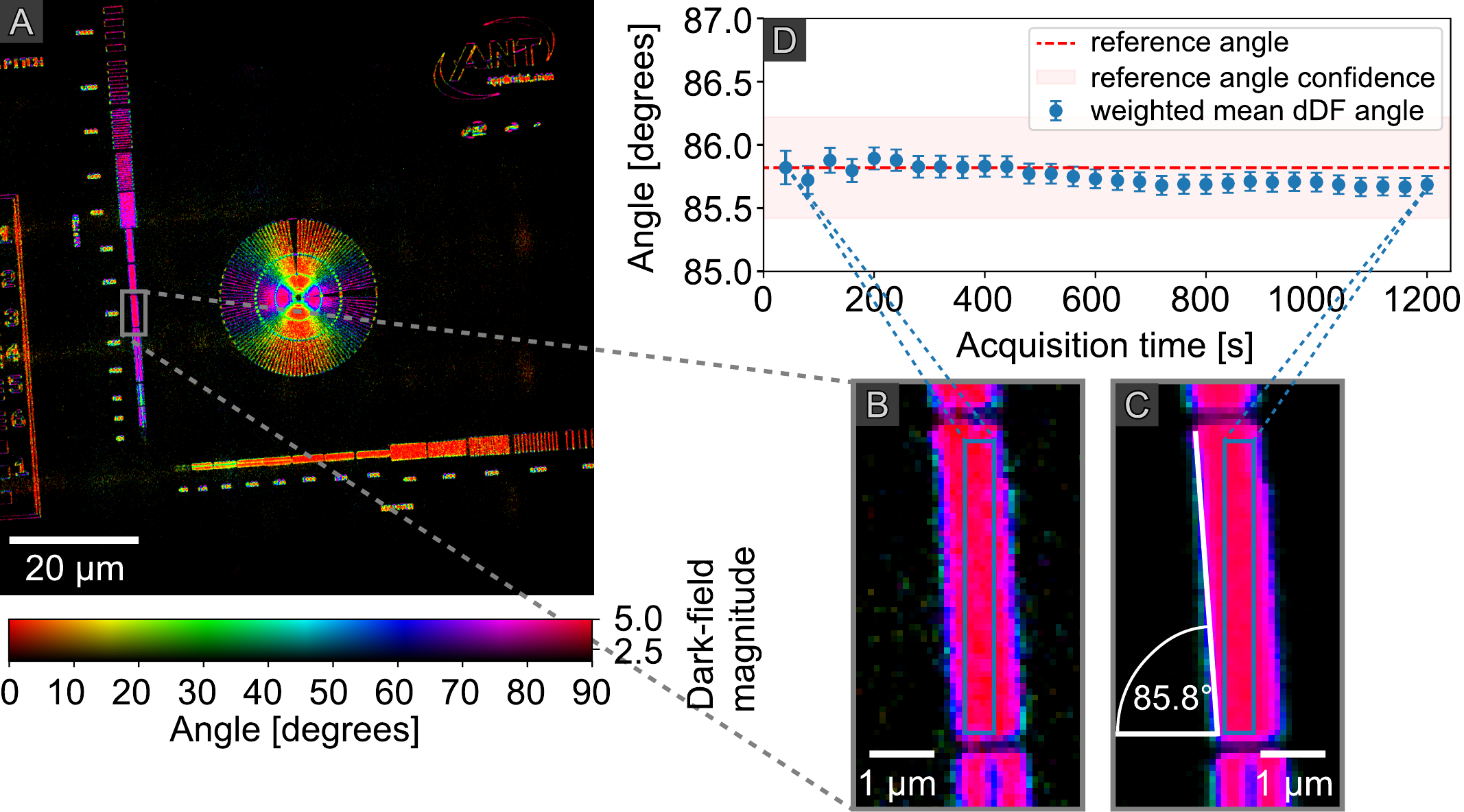}
    \caption{Noise evaluation of the directional dark-field image. The directionality can still be well separated with a total exposure time of $4 \times \SI{10}{\second} = \SI{40}{\second}$~(A). A zoom into the line pairs with a \SI{200}{\nano\meter} pitch (\SI{100}{\nano\meter} feature size)~(B) shows only minor differences to the same region exposed for \SI{20}{\minute}~(C). The magnitude-weighted mean of the directional dark-field angle for a region within the line pairs (marked by a blue rectangle in (B) and (C)) is calculated and plotted against the exposure time in (D). Furthermore, a reference angle of $\SI{85.8}{\degree} \pm \SI{0.4}{\degree}$ is measured manually to confirm the values of the directional dark-field angle and plotted as a red dashed line in (D).}
    \label{fig:statistical_analysis}
\end{figure}

While the measurements in Figure~\ref{fig:direc_df},~\ref{fig:si_direc_df}, and \ref{fig:tooth_direc_df} were acquired with long total exposure times, directional dark-field imaging remains robust even with significantly lower total exposure times. The statistical properties of the signal are examined in Figure~\ref{fig:statistical_analysis}. Figure~\ref{fig:statistical_analysis}~A shows the same Siemens star test pattern as in Figure~\ref{fig:direc_df} but with a total exposure time of \SI{40}{\second}. A zoom into the region of line pairs with a structure size of \SI{100}{\nano\meter} compares the signal of a measurement with a total exposure time of \SI{40}{\second} (Figure~\ref{fig:statistical_analysis}~B) to the signal from a measurement with a total exposure time of \SI{20}{\minute} (Figure~\ref{fig:statistical_analysis}~C). The comparison reveals only minor changes, mainly at the edges of the line pairs. A region inside the line pairs, marked by a blue rectangle, is used to calculate the magnitude-weighted mean angle and its standard deviation. The corresponding values are plotted over the total exposure times in Figure~\ref{fig:statistical_analysis}~D. Furthermore, a manually measured reference angle of $\SI{85.8}{\degree} \pm \SI{0.4}{\degree}$ is plotted as a red dotted horizontal line (measured \SI{90}{\degree} relative to the line pair orientation). The reference angle is also drawn in Figure~\ref{fig:statistical_analysis}~C. As visible in the plot of Figure~\ref{fig:statistical_analysis}~D, the magnitude-weighted mean angle shows only small variations throughout all total exposure times. Even at a total exposure time of \SI{40}{\second}, the magnitude-weighted mean exhibits a consistent value of $\SI{85.82}{\degree} \pm \SI{0.13}{\degree}$ with a small standard deviation. These results are strengthened by examining the signal-to-noise ratio of the magnitude for these regions, which is $28.49$ for the \SI{40}{\second} and $58.21$ for the \SI{20}{\minute} total exposure time.
These values suggest that the total exposure time can be further reduced without compromising the validity of the directional dark-field signal. However, the minimal feasible total exposure time depends on the signal strength, and hence on the structural properties of the examined specimen.

\subsection{Extension of the scattering vector}
\begin{figure}[htbp]
    \centering\includegraphics[width = \textwidth]{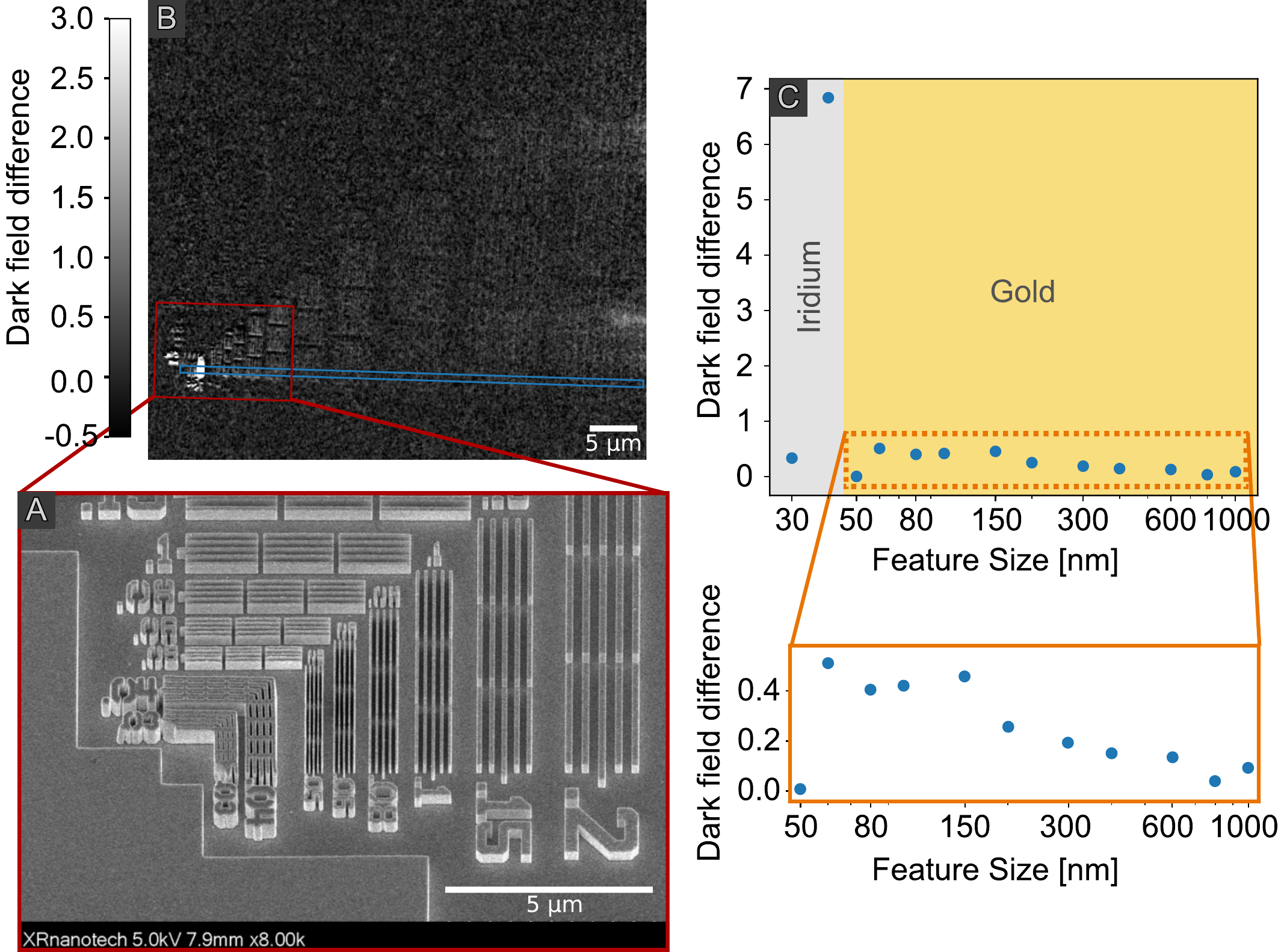}
    \caption{The dark-field signal is enhanced by utilizing the elongated shadow area of directional dark-field imaging. A SEM image shows the measured features (A). The dark-field image with closed condenser apertures (C-AP) is subtracted from the extended dark-field image with farther opened dark-field apertures (DF-AP) to visualize the additionally gained scattering intensity (B). The difference in the average dark-field intensity per feature size is plotted (C) for the region of interest indicated by the blue box in B. Note, in contrast to the rest, the feature sizes of \SI{40}{\nano\meter} and \SI{30}{\nano\meter} consist of iridium instead of gold and have a wall width to air gap ratio of 2:1, with the air gap being the main feature size. For the gold features, the ratio is 1:1. Each of the dark-field projections is exposed for \SI{200}{\second}.}
    \label{fig:ext_df}
\end{figure}

In the directional dark-field setup, the additional shadow originating from the C-AP can be utilized to push towards higher q-values and therefore smaller detectable feature sizes. This is investigated in a second experiment, where the extension of the scattering vector is examined using an elbow test pattern. A scanning electron microscope image of the test pattern is shown in Figure \ref{fig:ext_df} A. For feature sizes between \SI{1000}{\nano\meter} and \SI{50}{\nano\meter}, the line pairs are made out of gold with a structure height of \SI{516}{\nano\meter}. For the gold structures, the width of the wall and the width of the air gap between the line structures are identical. For feature sizes of \SI{40}{\nano\meter} and \SI{30}{\nano\meter} the line pairs are made out of iridium and have a height of \SI{670}{\nano\meter}. Here, the width of the walls is two times larger than the air gap in between. This means, that in case of the \SI{40}{\nano\meter} line pairs, the walls have a width of \SI{80}{\nano\meter} and an air gap of \SI{40}{\nano\meter}. Note, that the iridium structures also have bridges connecting the walls to increase stability.
For the X-ray measurement, all DF-APs were closed, as well as the left C-AP, so that two-thirds of the condenser is covered. In this configuration, a first direction-selective dark-field projection is taken. Afterward, the right DF-AP is opened to fully utilize the elongated shadow area in the back focal plane, and a second direction-selective dark-field projection is taken.
Based on that, two direction-selective dark-field projections are gained, one based on the standard scattering vector and one based on the extended scattering vector. The projections are subsequently aligned via FFT-convolution. To visualize the differences, the standard direction-selective dark-field projection is subtracted from the extended direction-selective dark-field projection, resulting in Figure \ref{fig:ext_df} B. In the Figure, all values above zero mean an increase in the total dark-field signal by the additional opening of the DF-AP. Based on a region of interest (shown in the figure by a blue rectangle), the average dark-field difference is calculated for each feature size and plotted in Figure \ref{fig:ext_df} C.
In the plot, it is visible that the dark-field signal can be boosted based on the extension of the scattering vector. This means that the Q-range, the range of scattering angles contributing to the dark-field signal (see equation~\eqref{eq:Q}), can be increased. It is also visible that smaller structures benefit more than larger structures. The extended dark-field signal seems to increase linearly towards smaller feature sizes and collapses nearly down to zero for a feature size of \SI{50}{\nano\meter}.
Table \ref{tab:DFTXM} shows the parameters of the experimental setup. By the additional opening of the DF-AP, the maximum of the scattering vector magnitude can be increased from \SI{0.0102}{\per\angstrom} to \SI{0.0126}{\per\angstrom}, which corresponds to a change of the smallest feature size from \SI{61.72}{\nano\meter} to \SI{50}{\nano\meter}. The calculated smallest structure size of the extended dark-field setup is equal to the point where the boost of the dark-field signal by the additional Q-range collapses in Figure \ref{fig:ext_df} C. The reason thereof is that the first peak of the \SI{50}{\nano\meter} structures in the reciprocal space probably still lies outside the extended Q-range and hence does not contribute. Since the main part of the scattered intensity is accumulated around these peaks, the intensity gained by the extension of the Q-range is insignificant for the \SI{50}{\nano\meter} feature size. The signal-to-noise ratio of the initial dark-field signal, prior to subtraction, is $15.4$ for the \SI{50}{\nano\meter} feature size. This is close to the signal-to-noise ratio of $17.0$ for the \SI{60}{\nano\meter} feature size, which shows the maximal signal gain. Hence, the lack of signal increase for the \SI{50}{\nano\meter} feature size is not attributable to detector noise, supporting the occurrence of reciprocal-space limits.
If we go towards smaller feature sizes of \SI{40}{\nano\meter} and \SI{30}{\nano\meter}, we see again an increase in the dark-field signal. This can be explained by the different design and material of these line pairs, as mentioned above.
For them, the feature size corresponds to the size of the air gap instead of the width of the material-filled part and has a different periodicity compared to the gold line pairs. 
Additionally, the smaller line pairs are made out of iridium and therefore have a higher electron density than the gold line pairs. Since in SAXS the scattering signal depends on the difference in electron density between the scatterer and the surrounding material, the signal is expected to be larger for iridium than for gold.  
These differences lead to a changed response of the dark-field signal, which could explain the stronger signal boost for the iridium line pairs compared to the gold line pairs.

\begin{table}
    \caption{Parameters for the experimental dark-field TXM setup. The spatial resolution is obtained by the 20\% criterion of the modulation transfer function of a Siemens star projection in transmission.}
    \setlength{\tabcolsep}{3pt}
    \centering
    \begin{tabular}{|c|c|c|c|c|c|c|c|}
        \hline
        spatial resolution & $r_{\text{C}}$ & $\Delta r_{\text{C}}$ & $r_{\text{B}}$ & $|\Vec{Q}_{\text{max}}|$ & $|\Vec{Q}_{\text{max, ext}}|$ & $d$ & $d_{\text{ext}}$\\
        \hline
        \SI{115}{\nano\meter} & \SI{0.9}{\milli\meter} & \SI{50}{\nano\meter} & \SI{0.55}{\milli\meter} & \SI{0.0102}{\per\angstrom} & \SI{0.0126}{\per\angstrom} & \SI{61.72}{\nano\meter} & \SI{50}{\nano\meter}\\
        \hline
    \end{tabular}
    \label{tab:DFTXM}
\end{table}

\section{Discussion}
\label{sec:Discussion}
\subsection{Directional dark field design decisions}
The capability of the presented approach to retrieve scattering orientations of nanostructures has been confirmed using a Siemens star test pattern (Figure~\ref{fig:direc_df}). Its robustness has further been demonstrated by cross-correlating orientational variations within hierarchical nanoporous silicon (Figure~\ref{fig:si_direc_df}), and its broader applicability has been illustrated by mapping the directional arrangement of hydroxyapatite nanocrystals in human tooth enamel (Figure~\ref{fig:tooth_direc_df}).

To retrieve the directionality of the dark field, additional apertures are placed upstream of the condenser, the C-AP. The apertures are like the DF-APs, placed on piezoelectric actuators, allowing fast and precise opening and closing. Due to the high precision of the piezos, the setup can seamlessly be switched between conventional transmission, dark-field, and directional dark-field imaging without realigning or refocusing. Blocking the beam upstream of the condenser prevents higher diffraction orders and unwanted scattering within the system, which could distort the image formation. Furthermore, the C-AP should not be placed downstream of the condenser, as the beamlets are diffracted towards the sample, where they superimpose. Consequently, the different fields begin to overlap at a point between the condenser and the sample. Due to the required thickness of the C-AP, the two L-shaped apertures can only be positioned at different distances from the condenser. The C-AP positioned closer to the sample interacts with more overlapping fields, which leads to ramp effects and an imbalance in the direction-selective dark-field projections. In the worst case, this could prevent the successful separation of the directional dark field.

The current implementation is based on four projections per directional dark-field image and is limited to an angular range of \SI{90}{\degree}. Four projections are needed since each projection suffers from a linear gradient in the bright field illumination. Currently, the condenser fields are arranged in circular patterns, but the C-AP is straight. By covering two-thirds of the condenser, the C-AP crosses the individual condenser fields with a different overlap and therefore blocks different area fractions of the individual fields. This leads to a bright field ramp in the detector plane, where the fields superimpose.
In the example of the C-AP being closed at the top, a bright-field gradient runs from the top (dark) to the bottom (bright). This unbalanced flux leads to a different dark-field intensity at the bottom and the top part of the image. Since the projection with the bottom C-AP closed suffers from the same gradient but running from the bottom (dark) to the top (bright), two projections can be combined to cancel out the gradient. The combined direction-selective dark-field image becomes flat again (Figure \ref{fig:direc_df} A and B). So, per direction, two projections are needed to achieve a flat dark-field response. 
To reduce the impact of the bright-field ramps, the shape of the condenser or the C-AP needs to be adapted accordingly.

If the illumination ramp in a single dark-field projection with the C-AP closed can be corrected, the directionality could be extended to more than \SI{90}{\degree}. Due to the necessity of combining two projections, the sensitivity towards positive and negative direction is lost. This reduces the angular range from \SI{360}{\degree} to \SI{180}{\degree} for the x-direction, and since the same is true for the y-direction, the angular range is further decreased to \SI{90}{\degree}. Note that while the setup could potentially differentiate the scattering direction to an angular range of \SI{360}{\degree}, this does not mean that the setup is able to break the symmetrical properties of the scattering signal. The setup can therefore still be limited to smaller angular ranges depending on the signal itself.

\begin{figure}[htbp]
    \centering\includegraphics[width = \textwidth]{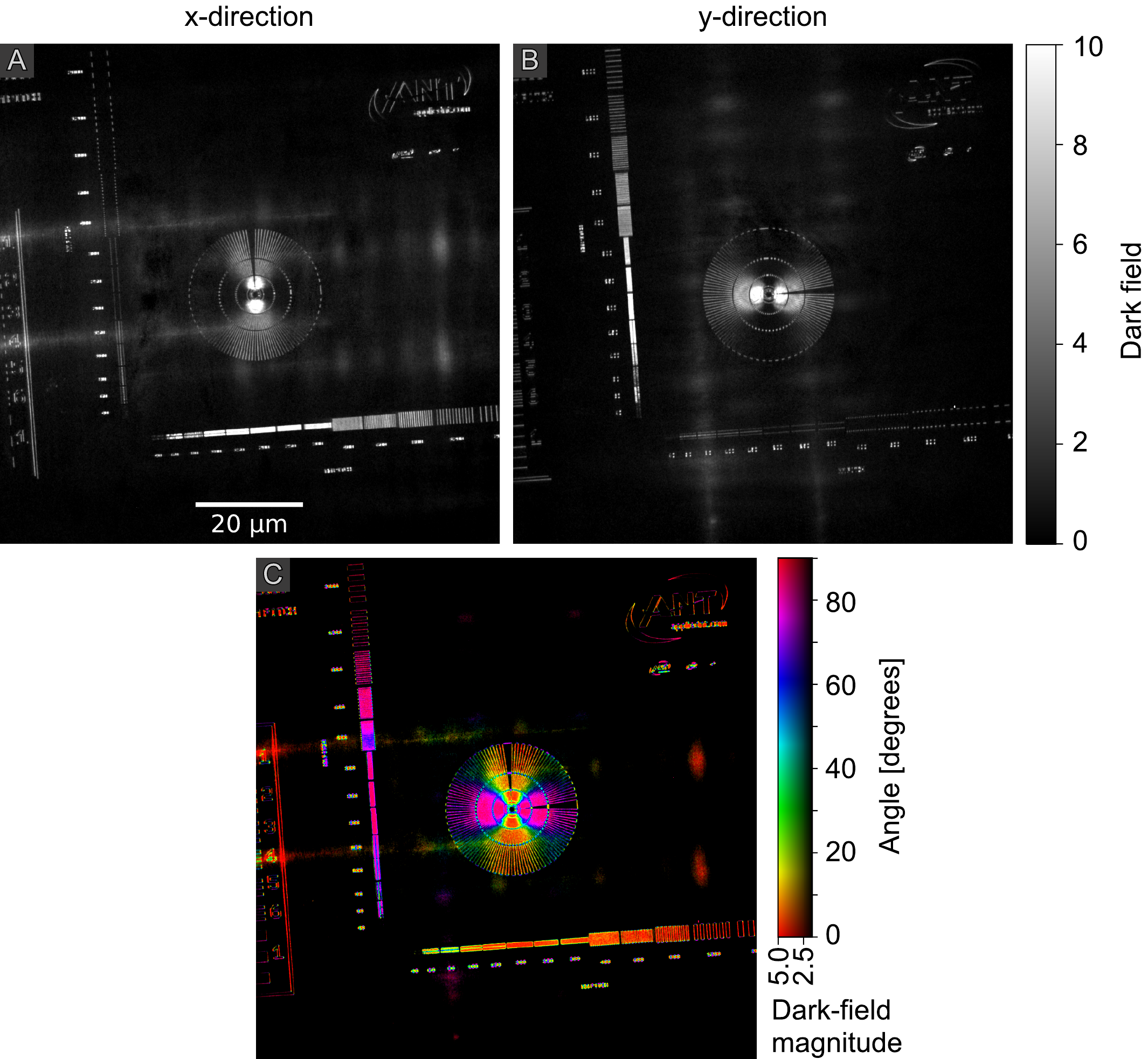}
    \caption{Directional dark field of a Siemens star acquired with the slit system. The condenser apertures (C-AP) are closed to form once a horizontal and once a vertical slit, resulting in the direction-selective dark-field image in the x-direction (A) and the y-direction (B). Based on these components, a scattering vector is calculated (C). The color corresponds to the scattering angle, and the luminance to the magnitude of the scattering vector. The background artifacts also become visible in the directional dark-field image, leading to locally distorted angles. The total exposure time of the directional dark-field image is $2 \times \SI{300}{\second} = \SI{10}{\minute}$.}
    \label{fig:direc_df_slits}
\end{figure}

The direction-selective dark-field projections can alternatively be obtained directly by closing the C-AP once to a horizontal slit (scattering in x-direction) and once to a vertical slit (scattering in y-direction). This allows to extract directly $D_{\text{x}}$ and $D_{\text{y}}$ in equation~\eqref{eq:Dx} and \eqref{eq:Dy}. The corresponding retrieved direction-selective dark-field projections are shown in Figure \ref{fig:direc_df_slits}~A and~B. By comparing the projections to Figure \ref{fig:direc_df}~A and~B, it becomes visible that some artifacts start to occur. The artifacts lead to an offset in the local dark-field values and therefore distort the calculated angle, as seen in Figure \ref{fig:direc_df_slits}~C, and are challenging to correct. This is especially prominent in the horizontally oriented line pairs on the left side. Since these artifacts remain even after the subtraction of the background scattered intensity $I_{\text{b}}$ in equation~\ref{eq:Dr}, the artifacts must be connected to the sample itself. We hence believe that these are some diffraction or reflection effects from the C-AP, which interact with the sample, sample membrane, or sample pin and scatter back into the field of view.
Furthermore, as only two projections are taken, the angular sensitivity of the slit approach is limited to \SI{90}{\degree} and only the absolute scattered intensity per projection is measured. Scattering components in positive and negative directions cannot be differentiated, as discussed above. In terms of total exposure times, both approaches are comparable. For the slit approach, two-thirds of the condenser (upper and lower thirds) are covered. However, due to the round shape of the condenser, the number of condenser fields in the central third is lower. It contains 24 fields compared to 34 fields in each of the top and bottom thirds. Hence, the slit approach requires $1.41$ times the total exposure time.

We demonstrated that directional dark-field projections can be performed with a total exposure time of \SI{40}{\second}, achieving high angular accuracy and a magnitude signal-to-noise ratio of $28.49$. This suggests that the total exposure time can be lowered without compromising the validity of the signal. By optimizing the optics, particularly the shape of the condenser, the beamstop, and the apertures, the flux within the system can be enhanced, and the required total exposure time can be further reduced. Additionally, the experiments shown here were conducted at a third-generation synchrotron. The advantages of a fourth-generation synchrotron will enable a significant reduction of the total exposure time, which could ultimately allow time-sensitive in-situ experiments. Current possible applications could include stress deformation in fibril samples, such as those found in bone, teeth, and carbon composites, or other hierarchical nanoporous materials, as well as their reaction to environmental changes, including temperature and humidity.

Highly orientation-dependent nanostructures cause strong anisotropic scattering signals that can be visualized in directional dark-field imaging. But since the scattering signal depends on the orientation of the structures relative to the beam, well-structured features such as fibers break the rotational invariance, causing problems using conventional reconstruction algorithms for computed tomography \cite{Liebi2018, Nielsen2024, Lautizi2024}. By adding a second rotation axis perpendicular to the beam direction, the nanostructures can be recorded from virtually all possible sample orientations to reconstruct the full three-dimensional tensor in each voxel \cite{Lautizi2024}. Hence, the proposed method can be extended in the future to tensor tomography by adding a second rotation axis to extract the three-dimensional feature orientation.

\subsection{Extension of the scattering vector}
By utilizing the extended shadow region of the closed C-AP, we were able to increase the maximum magnitude of the scattering vector from \SI{0.0102}{\per\angstrom} to \SI{0.0126}{\per\angstrom} and thus enhance the signal, particularly for smaller feature sizes (Figure \ref{fig:ext_df}). The signal increase dropped for a gold feature size of \SI{50}{\nano\meter}, corresponding to the derived limit of the extended dark-field setup. However, inconsistencies in the condenser and FZP, such as collapsed zones, can lead to a deviation from the theoretically calculated limit. Furthermore, the theoretical limit does not necessarily correspond to feature sizes that are no longer visible in the dark field. For these feature sizes, the first peak in the reciprocal space lies outside the q-range of the setup, but can still contribute to the image formation, as discussed in Wirtensohn et al. \cite{Wirtensohn2024}. Either way, the experiment shows that the response to different feature sizes can be controlled by modifying the illumination and adapting the DF-APs. With a well controlled illumination pattern and matched apertures, this could ultimately lead to size-selective dark-field imaging, where the feature sizes contributing to the dark-field signal can be tuned precisely.

\section{Conclusion}
Recent approaches show promising results utilizing the directional dependence of the dark-field signal in projection space. However, these methods are limited to the micrometre range.
Here, the first full-field directional dark-field TXM setup for nano imaging is proposed, allowing the orientation of anisotropic sample structures to be extracted. The successful retrieval of the scattering orientation for the spikes of a Siemens star test pattern, the orientation decomposition of line pairs, and the extraction of the average orientation within one pixel for feature sizes below the spatial resolution of the setup are demonstrated. This provides information beyond conventional dark-field, attenuation, and Zernike phase-contrast imaging. Based on a scan of hierarchical nanoporous silicon and human tooth enamel, the applicability of this approach to samples in the fields of biomedicine and materials science is shown. Furthermore, the Q-range is extended by manipulating the illumination function, increasing the detectability of smaller features and is a first step towards size-selective dark-field imaging. The setup only requires an additional aperture, allowing existing dark-field TXMs to be extended to directional dark-field imaging with little effort.

\begin{backmatter}

\bmsection{Funding} The authors gratefully acknowledge financial support by the ERC consolidator grant (Julia Herzen, TUM, DEPICT, PE3, 101125761) and by the ECI pathfinder (1MICRON, 101186826). 
Furthermore, this work was supported by the Deutsche Forschungsgemeinschaft (DFG) within the Collaborative Research Initiative CRC 986 “Tailor-Made Multi-Scale Materials Systems” project number 192346071, CRC 1615 "SMART REACTORS -
Reactors for Future Process Engineering" project number 503850735, and as part of the Excellence Strategy of the Federal Government and the federal states -- BlueMat: Water-Driven Materials -- EXC 3120/1 -- 533771286.

\bmsection{Disclosures} The authors declare no conflicts of interest.

\bmsection{Data availability} Data underlying the results presented in this paper are not publicly available at this time but may be obtained from the authors upon reasonable request.

\bmsection{Acknowledgments}
We acknowledge Helmholtz-Zentrum Hereon (Geesthacht, Germany) and DESY (Hamburg, Germany) for providing the experimental facilities. This research was carried out at PETRA~III. Beamtime was allocated for proposal I-20240214.
This research was supported in part through the Maxwell computational resources operated at Deutsches Elektronen-Synchrotron DESY, Hamburg, Germany. The study of the human tooth was reviewed and approved by the Ethics Committee of the Medical University of Vienna under approval number 1065/2013. Furthermore, we thank Gudrun Lotze for her comments.

\bmsection{Author contributions}
S.W. wrote the main manuscript.
S.W., S.F., and I.G. conducted the experiments.
S.W., D.J., S.F., and I.G. finalized the manuscript.
S.W., S.F., I.G., D.J., and J.H. conceptualized the manuscript.
P.H., M.M., D.H., and I.K. developed and provided the Si sample, its SEM images, and the related materials science context.
K.B., S.T., and C.K. created and provided the tooth sample as well as the related biomedical context.
P.Q. and C.D. created and provided the optics.
S.F. and I.G. supervised the project.
K.S. prepared the samples.
All authors reviewed the manuscript.

\end{backmatter}


\bibliography{sample}

\newpage
\section{Supplements}
\begin{figure}[htbp]
    \centering\includegraphics[width = \textwidth]{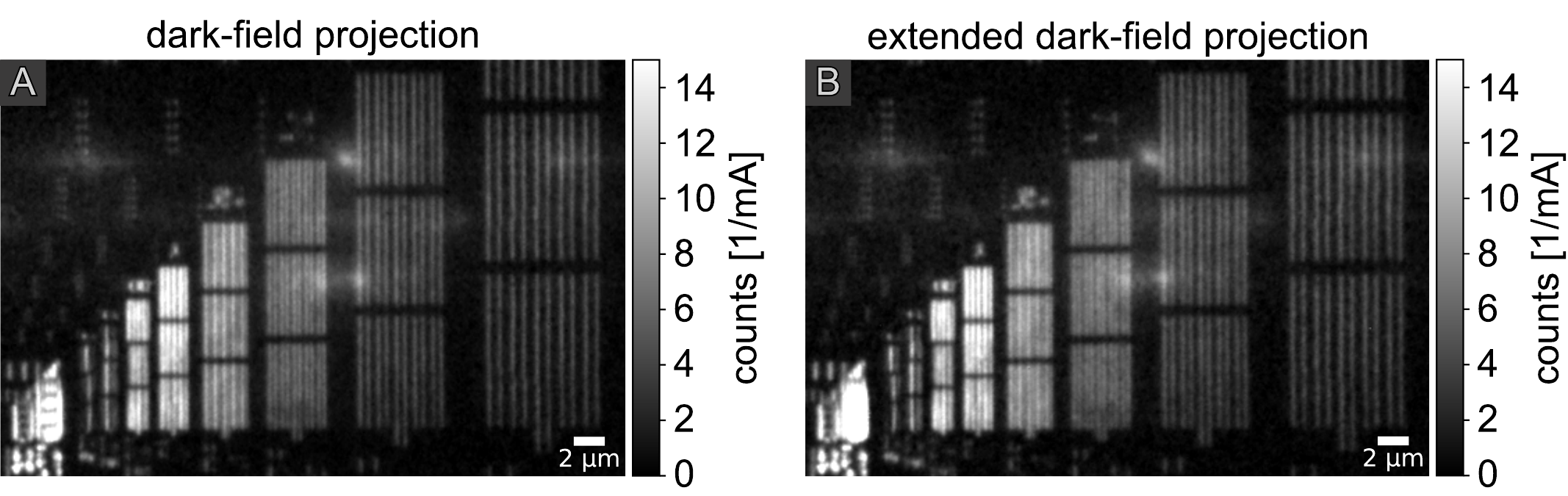}
    \caption{Dark-field projection of a line pair test pattern with the left CAP closed (A). Due to the elongated shadow in the back focal plane of the FZP, the right DF-AP can be further opened to accept a larger maximum magnitude of the scattering vector, increasing the intensity especially for smaller feature sizes (B). The difference of these dark-field projections is used in Figure~\ref{fig:df_extension}. Each of the dark-field projections is exposed for a total of \SI{200}{\second}.}
    \label{fig:proj_ext_df}
\end{figure}

To gain the difference image in Figure~\ref{fig:df_extension}~B, the dark-field projection shown in Figure~\ref{fig:proj_ext_df}~A is subtracted from the extended dark-field projection in Figure~\ref{fig:proj_ext_df}~B. Due to the high (dynamic) range of the dark-field intensity values, the change is barely visible in the direct comparison.

\end{document}